\documentclass[12pt,preprint]{aastex}

\tighten

\slugcomment{accepted for the publications in the Astrophysical Journal  on November 14, 2010}
\def\lsim{\lower.5ex\hbox{$\; \buildrel < \over \sim \;$}}
\def\gsim{\lower.5ex\hbox{$\; \buildrel > \over \sim \;$}}
\def\lax {\ifmmode{_<\atop^{\sim}}\else{${_<\atop^{\sim}}$}\fi}
\def\gax {\ifmmode{_>\atop^{\sim}}\else{${_>\atop^{\sim}}$}\fi}

\def\gtorder{\mathrel{\raise.3ex\hbox{$>$}\mkern-14mu
\lower0.6ex\hbox{$\sim$}}}
\def\ltorder{\mathrel{\raise.3ex\hbox{$<$}\mkern-14mu
\lower0.6ex\hbox{$\sim$}}}

\def\pmb#1{\setbox0=\hbox{#1}%
\kern-0.015em\copy0\kern-\wd0
\kern0.03em\copy0\kern-\wd0
\kern-0.015em\raise0.0433em\box0 }

\begin{document}

\title{Spectral Index as a Function of Mass Accretion Rate  in Black Hole Sources.
Monte-Carlo Simulations and an Analytical Description}

\author{Philippe Laurent\altaffilmark{1} and Lev Titarchuk\altaffilmark{2,3,4} }

\altaffiltext{1}{CEA/DSM/IRFU/APC, CEA Saclay, 91191 Gif sur Yvvete, France;
plaurent@cea.fr}

\altaffiltext{2}{Physics Department, University of Ferrara,
Via Saragat, 1 44100 Ferrara, Italy; titarchuk@fe.infn.fe}

\altaffiltext{3}{Goddard Space Flight Center, NASA,
Astrophysics Science Division, code 663, Greenbelt MD 20770, USA
lev@milkyway.gsfc.nasa.gov}

\altaffiltext{4}{George Mason University  Fairfax, VA 22030, USA}

\vskip 0.5 truecm


\begin{abstract}
We present herein a theoretical study of correlations between spectral indexes of X-ray emergent spectra and mass accretion rate ($\dot m$) in black hole (BH) sources, which provide a definitive signature for BHs. 
It has been firmly established, using the {\it Rossi X-ray Timing Explorer} ({\it RXTE}) in numerous BH observations during hard-soft  state spectral evolution, that the photon index of X-ray spectra  increases when $\dot m$ increases and, moreover, the index saturates at  high values of  $\dot m$.
In this Paper, we present  theoretical arguments  that the observationally established index saturation effect  vs mass accretion rate
is  a  signature of the bulk  (converging) flow onto the black hole. Also, we demonstrate that the index saturation value  depends on the plasma temperature of converging flow.
We self-consistently calculate the Compton cloud (CC) plasma temperature as a function of mass accretion rate using the energy balance between energy dissipation and Compton cooling.  We explain the observable phenomenon, index-$\dot m$ correlations using   a Monte-Carlo  simulation of   radiative processes in the innermost part (CC) of  a BH source and we account for the Comptonization processes in the presence of thermal and bulk  motions, as  basic types of plasma motion.
We show that, when $\dot m$ increases,  BH sources  evolve to high and very soft states  (HSS and VSS, respectively), in which  the strong blackbody-like  and steep power-law components   are formed  in the resulting X-ray spectrum. The simultaneous detections of these two components strongly depends on sensitivity of high energy instruments,  given that the relative contribution of the hard power-law tail in the resulting VSS  spectrum can be very low, which is why, to date {\it RXTE} observations of the VSS  X-ray spectrum has been characterized by the presence of the strong BB-like component only.
We also predict specific patterns for high-energy efold  (cutoff) energy ($E_{fold}$) evolution with $\dot m$ for thermal and dynamical (bulk) Comptonization cases. For  the former case,  $E_{fold }$ monotonically decreases with $\dot m$, in the latter case,  the $E_{fold}-$decrease is followed by its increase at high values of  $\dot m$. The observational evolution of $E_{fold}$ vs $\dot m$ can be one more test  for the presence of a converging flow effect in the formation of the resulting spectra in the close vicinity of BHs.
 \end{abstract}

\keywords{black hole physics---accretion disks --- radiation mechanisms: nonthermal---X-rays: general}

\section{Introduction}
 A study of the characteristic changes in spectral and variability properties  of X-ray binaries has proven to be a valuable source of information on the Physics
governing the accretion processes and on the fundamental parameters of  black hole  (BH) sources.  BH observational appearance is conventionally described in terms of BH state classification \citep[see][for different definitions of BH states]{rm,bell05,kw08}.
We adopt the following BH state classification for five major BH states: the {\it quiescent}, {\it low-hard} (LHS), {\it intermediate} (IS), {\it high-soft} (HSS) and {\it very soft} states (VSS). When a BH transient  goes into outburst, it leaves the quiescent state and enters  the LHS, a low luminosity state with the energy spectrum dominated by a Comptonization component
combined (convolved) with  a  weak thermal component. The photon spectrum in the LHS
is presumably a result of  Comptonization (upscattering) of soft  photons, originating  from a relatively weak accretion disk,  off electrons of  the hot ambient plasma [see e.g. \cite{ST80}, hereafter ST80].
 The HSS photon spectrum is characterized by a prominent thermal
component which is probably a signature of  a strong emission coming from a geometrically thin accretion disk. A weak power-law component is also present at the level of not more than 20\% of the total source flux.

Timing and spectral properties of an accreting BH are tightly
correlated for a number of BH sources [see \cite{vg03} and  a comprehensive study of these  correlations was done by  \cite{st09},  \cite{ts09},  hereafter ST09 and TS09, respectively].
Correlations between the spectral hardness (photon index) and
the characteristic frequencies of the quasi-periodic oscillations (QPOs)
observed in the lightcurves of BH   sources has been proposed to use as a tool
to determine a BH mass \citep[][hereafter TF04]{tf04}.
\cite{rm} demonstrated and then ST09 confirmed that there is a  correlation between values QPO frequencies and the disk flux, namely mass accretion rate $\dot M$ in the disk, for a number of  BH candidates (BHCs).  Moreover, ST09, TS09 and \cite{st10}, hereafter ST10, also found  strong correlations between spectral index and $\dot m$, where $\dot m= \dot M/\dot M_{Edd}$ is  dimensionless mass accretion rate in units of critical mass accretion rate $\dot M_{Edd}=L_{Edd}/c^2$,   in 19 spectral transition episodes from 10 BH sources observed  with the {\it Rossi X-ray Timing Explorer (RXTE)}.
ST09  combine these index-QPO and index-$\dot m$ correlations  to measure  BH masses  in 7 BH sources.  For some of these sources, (i.e. H 1743-322, Cyg X-1, XTE 1650-500, GX 339-4 and XTE J1859-226) this new ST09 method provides  a much better precision than conventional dynamical methods (see  ST09). In addition TS09 and ST10 using  index-$\dot m$ correlations  evaluate BH masses in GRS 1915+105  and SS 433 respectively.

The index-QPO and index-$\dot m$ correlation has a specific shape characterized by a rise part followed by the saturation plateau.   The ST09  method  is based on scaling (sliding) the correlation pattern related to  a reference  source vs that of a given source for which a BH mass is determined.
This determination can be only done for this pair of the sources  if the saturation values  of index and inclination of the rising part are the same for both.  Thus, the correlation patterns need to be self-similar to implement this scaling and, finally, to obtain a BH mass for a given source.
 Using results from \cite{LT99}, hereafter LT99, TF04 who were the first to suggest a model for the index-$\dot m$ correlation observed in BHCs, we further explore the model of  index-$\dot m$ correlation  and present  a thorough  modeling  of this correlation using Monte Carlo simulations for a wide range of basic model parameters.
We also investigate the possibility that the shape of the correlation
pattern provides a direct signature of the bulk motion (converging) flow onto  a black hole,  which {\it would} be the signature of the black hole [see \cite{tmk97},  \cite{tz98} and LT99 for more details on this subject]. We present theoretical arguments based on the Monte-Carlo simulation and analytic consideration that
 the index-mass accretion rate saturation effect observed in a number of BHCs is a signature of a BH converging flow. This saturation effect should occur  when the  mass accretion rate
 $\dot M$ exceeds the Eddington limit $\dot M_{\rm Edd}$, which can only  exist in BH sources, see \cite{tmk97}.

The origin of the spectral state transition and its final end as a steep power-law in the high soft states is still debated in the literature [see e.g. \cite{ct95}, hereafter CT95,  LT99, \cite{lt01}, \cite{nz06} and \cite{rm}, hereafter RM06]. However, there is  agreement within the community that the spectral transition is driven by mass accretion in the system (see e.g. CT95, RM06) such that, when $\dot m$ increases a BH source goes from the LHS to soft states (HSS or VSS) through the intermediate state (IS), but a detailed scenario for this transition is still missing.   A natural question to ask is: why does the source go to either the HSS  or the VSS and what physical processes are behind of each of these transitions?
The soft component of the observed X-ray spectra is usually  fitted by a blackbody (BB) shape modified by Comptonization for which the  Comptonization parameter $Y_{sf}$ decreases, or the energy index of the Comptonization Green's function  $\alpha_{sf}$ increases towards a softer state [see details in TS09].  Furthermore, when the mass accretion rate increases up to some critical value photon index $\Gamma_{sf}=\alpha_{sf}+1$ starts to saturate to a value of about 4.2. What mechanism is responsible for the behavior of $\Gamma_{sf}$ vs $\dot m$ and for the saturation level of the index? In contrast, ST09  and TS09 show that the index of the hard component of the X-ray spectrum $\Gamma_h$ saturates to different levels  that varies between 2 to 3 for different sources and for different outbursts in a given source. For example, $\Gamma_h$ saturates to 3 in GRS 1915+105 (TS09) and to 2.3 in SS 433 (ST10) and it varies from 2 to 2.6 for different outbursts in  GX 339-4  (see ST09).

Another phenomenon revealed in X-ray observations of compact objects
[neutron stars (NSs)  and BHs]  is related to the evolution of a high energy  efold (cutoff) energy $E_{fold}$ of the spectra during the spectral evolution. For NSs, $E_{fold }$ steadily decreases from the hard state to  the soft state
[\cite{ts05}, \cite{ft10}] but, in BHs,  the observational pattern of $E_{fold}$ vs $\dot m$ is such that   $E_{fold }$ begins to decrease with $\dot m$, which is then followed by a shift of $E_{fold}$ to higher values when $\dot m$ increases \citep{ts10}.

 We present our picture of accretion in the innermost part of source, the Comptonization region [Compton cloud (CC)]  and a model to calculate the CC plasma temperature
  in \S \ref{model}. We provide  the details of the Monte-Carlo (MC) simulations in \S \ref{MC}.
  We present a combined effect of thermal and bulk motion Comptonization and BH spectral evolution in
  \S\ref{thermal+bulk}.
We show and discuss our MC simulation results and their theoretical explanation related to the index-mass accretion rate correlation in \S \ref{MC_index-mdot}.  Specifically,
 we discuss the signature of bulk motion Comptonization (BMC) and its relation to the index evolution during state transition.   Also, we show that the index saturation
effect is a direct consequence of the existence of the innermost  bulk motion
region and, therefore, can be considered to be an observational signature
of a BH.
 $E_{fold}$-$\dot m$ correlations results and their interpretation are shown in \S \ref{E_cf-mdot}.  Conclusions   follow in \S \ref{summary}.

\section{The model}\label{model}
\begin{figure}[ptbptbptb]
\includegraphics[scale=1.,angle=0]{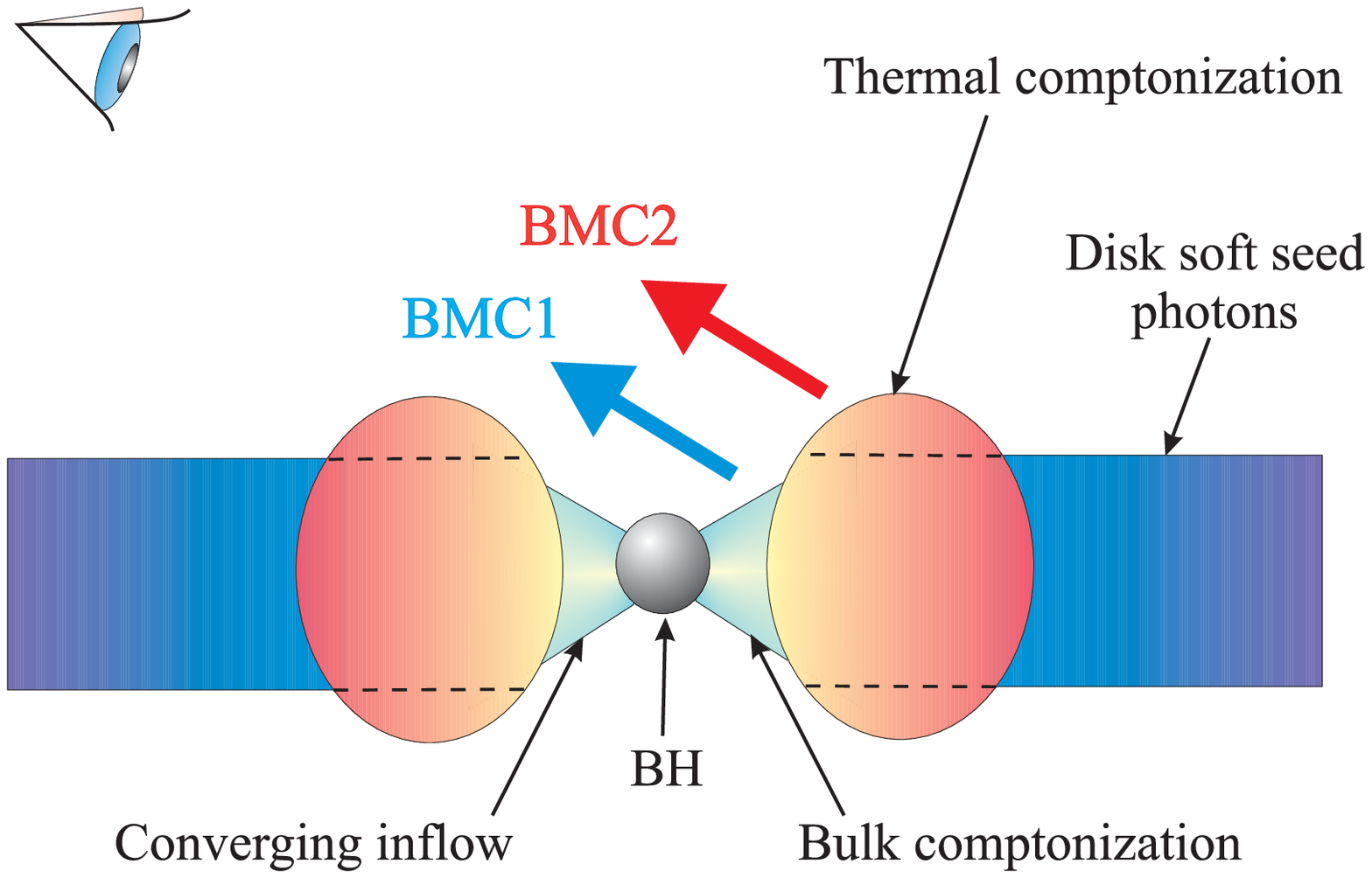}
\caption{A schematic view of the proposed geometry for thermal and bulk Comptonization regions in a source hosting a BH with power law-like emission at high energies.  The bulk Comptonization  plus thermal spectrum (bulk {\it BMC1} plus thermal {\it BMC2}) arises in the innermost part of the transition layer (TL), where the disk blackbody-like  seed photons are (thermally and
dynamically) Comptonized by the in-falling material. 
}
\label{geometry}
\end{figure}

We illustrate our accretion scenario in Figure \ref{geometry}, which we suggest  taking  place in the innermost part  of a black hole source. We assume that an accretion flow onto a black hole (BH)  consists of three parts: a geometrically thin  accretion  disk
[standard Shakura-Sunyaev disk, see \cite{ss73}], a transition layer (TL),  which is an intermediate link between the accretion disk, and a converging (bulk) flow region  (see TF04), that is assumed to exist, at least, below 3 Schwarzschild radii, $3r_{\rm S}=6GM_{bh}/c^2$.
There is substantial observational evidence for this  three  component model for accretion flows onto black holes [see \cite{tsa07}, hereafter TSA07, ST09, TS09 and ST10]. In addition, power density spectra (PDSs) during the transition from the low-hard to high-soft states indicate low and high frequency white-red noise (WRN) components, with discrete Lorentzian QPO features superposed. The low frequency
WRN component dominates in the high-soft state PDSs while the high frequency  PDS component
dominates in  the low-hard state, with the interpretation that these two components represent
the diffusion of perturbations in an outer accretion disk and an inner Compton corona (transition layer), respectively. These two regions are probably separated by an accretion shock, which moves to smaller radii in the high-soft state and to larger radii in the low-hard state (TSA07).
This accretion shock represents a location
in the disk where the outer Keplerian flow begins to transition to the inner sub-Keplerian
motion of the compact object and its inner corona.

We assume that the plasma temperature of the converging flow  is near that of the TL (or corona).
The TL temperature can be presented in the analytical form using   Eq. (12)  in TF04\footnote{
 There is a typo in formula (13) of  TF04 which  is the solution of Eq. (12) in TF04.}:
\begin{equation}
\frac{T_e}{10^8{\rm K}}=F(\tau_0, q)=\frac{15}{\tau_0(1+3\tau_0/8)(1+q)-0.62}
\label{temp_tau_0}
 \end{equation}
where $\tau_0$ is TL optical depth  and $q=Q_d/Q_{TL}$ is a ratio of
the energy releases in the disk and transition layer, respectively.

 We relate $\tau_0$ to   optical depth  of the converging flow $\tau_{ff}$ region  by using the flow continuity equation
\begin{equation}
\tau_{0}\sim\frac{V_{ff}}{<V_{MA}>}|_{r=r_{out}}\tau_{ff},
\label{t_0-t_ff}
\end{equation}
where
the free-fall velocity
\begin{equation}
V_{ff}=c \,(r_{\rm S}/r)^{1/2},
\label{v_ff}
\end{equation}
$c$ is the speed of light, $r_{out}$  is an outer radius of the converging flow and $<V_{MA}>$ is a TL magneto-acoustic velocity  averaged over  the layer   $<V_{MA}>$ and approximately equals to  radial velocity of the TL flow.  
 
 The optical depth of the converging flow,  as evaluated by an observer at rest, is given by
\begin{equation}
\tau_{ff}=\dot m(\pi/2-\arcsin{x_{out}^{-1/2}}),
\label{t_ff}
\end{equation}
where $\dot m=\dot M/\dot M_{\rm Ed}$ is a dimensionless mass accretion rate in units of the critical mass accretion rate $\dot M_{\rm Ed}=L_{Ed}/c^2$ and $x_{out}=r_{out}/r_{\rm S}$ is the dimensionless outer radius of the converging flow in units of Schwarzschild radius 
$r_{\rm S}$. Note,  if the dimensionless BH spin $a$  is less than 0.8 then  $2.7<x_{out}\leq 3$ [see \cite{ll75}]. 

\cite{tbw01}, hereafter TBW01,  estimated the {\it B}-field around neutron stars (NS) based on observed kilohertz  and viscous quasi-periodic  oscillation  (QPO) frequencies.  They found that the best-fit values of $V_{MA}$ in the transition layer  related to  the {\it B}-field of NS   and plasma density is in the range of $10^8-10^9$ cm s$^{-1}$, 
where \\$V_{MA}\gax 10^9 (B/10^{6}~{\rm G})/(4\pi \rho/10^{-6}{\rm g~cm^{-3}})^{1/2}$ cm s$^{-1}$
 and $\rho$ is TL plasma density which is of order of $\rho=n_em_p= 1.6\times 10^{-5}(n_e/10^{19}{\rm cm}^{-3})$ g cm$^{-3}$.   
Unfortunately, there is no estimate of $V_{MA}$ for the innermost part of a BH that is based on  observations.   
\cite{tlm98}  introduced the Reynolds number   
$\rm Re= V_{R}R/\hat\nu$  ($\gamma$ in their notation)  where $V_{R}$,  $\hat\nu$  are an average radial velocity,  an average  viscosity  over a given configuration, respectively  and $R$ is a configuration scale.   They demonstrated that the size of the transition layer (CC)  between the fast rotating accretion disk and the relatively slow rotating central object (either BH or NS)  strongly depends on the Reynolds number $\rm Re$.  Moreover, \cite{ts08} infer, by analyzing X-ray data for Cyg X-1, that the Reynolds number 
$\rm Re$ of accretion flow in the transition layer increases from 10 to 100 when the source evolves from the low/hard state  to the high/soft state. This is possible when an increase of $V_R$ ($\sim V_{MA}$) leads  to  the $\rm Re$ rise. 

Because $\dot m$ is proportional to $V_R$ (and $\sim V_{MA}$) indicates  that 
$V_{MA}$ is also proportional to $\tau_{ff}$  (see Eq. \ref{t_ff} for linear relation between $\dot m$  and $\tau_{ff}$): 
\begin{equation}
<V_{MA}>=C_0+C_1\tau_{ff}
\label{v_ma-t_ff}
\end{equation}
where $C_0\sim 0.01c$ and $C_1/C_0\sim 0-10$.

The combination of Eqs. (\ref{t_0-t_ff}), (\ref{v_ff}) and (\ref{v_ma-t_ff}) gives $\tau_0$ as a function of $\tau_{ff}$ and $x_{out}$:
\begin{equation}
\tau_0\sim\frac{(c/C_0)\tau_{ff}}{[1+(C_1/C_0)\tau_{ff}]x_{out}^{1/2}}.
\label{t_0-t_ff_new}
\end{equation}

\begin{figure}[ptbptbptb]
\includegraphics[scale=0.7,angle=0]{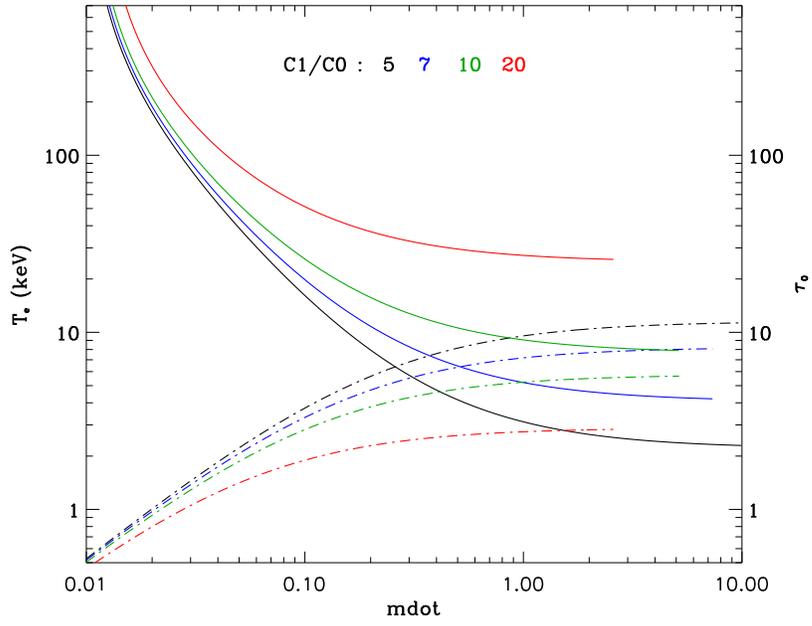}
\caption{Plasma temperature (solid lines) and optical depth (dash lines) of the transition layer   vs  $\dot m$.}
\label{fig2}
\end{figure}

Equations  (\ref{t_ff})  and (\ref{t_0-t_ff_new}) yield, for nonzero $C_1$,  TL optical depth
$\tau_0$  saturates to a constant value, $\tau_{0, sat}=c/(C_1x_{out}^{1/2})$ when $\tau_{ff}$ (or $\dot m$) increases.
This saturation of $\tau_0$ leads to the saturation of temperature $T_e$ with $\dot m$
for a given  $q$ (see Eq. \ref{temp_tau_0}). In this case, the temperature saturation value
(see Eq. \ref{temp_tau_0}) is
\begin{equation}
T_{e, sat} =F(\tau_{0,sat}, q)10^8{\rm K}.
\label{temp_sat1}
\end{equation}
The thermal Componization  index $\alpha_{sf}$ also saturates when $\tau_0$ and $T_e$ saturate given that  $\alpha_{sf}$ is a specific function of $\tau_0$ and $T_e$ [see e.g. \cite{ST80}, hereafter ST80].
In Figure \ref{fig2}, we present  $\tau_0$ and $T_e$ (in units of keV )  as functions of $\dot m$  calculated using Eqs. (\ref{temp_tau_0}) and  (\ref{t_0-t_ff_new}). These calculations are made for  $q=0.01$, but for all $q\ll1$ the results are similar to those presented in Figure  \ref{fig2}. It is clearly evident in Figure \ref{fig2} that  saturation of $\tau_0$ and $T_e$
occurs at high values of $\dot m$.

Using Eq. (\ref{temp_tau_0}) and for $\tau_0\gg1$,  the product
$T_e\tau_0^2\rightarrow 4\times10^9/(1+q)$ K.
The thermal Comptonization parameter
$Y_{sf} \propto (kT_e/m_ec^2)\tau_0^2$ (ST80), which implies that  $\alpha_{sf}\sim1/Y_{sf}$  goes to a constant too.
In \S \ref{thermal+bulk}, we demonstrate that this saturation effect of the thermal Comptonization index
is reproduced in our Monte Carlo simulations.  It is also worthwhile to emphasize that
the index saturation of the soft component of X-ray spectra  is detected in {\it RXTE} observations
of GRS 1915+105 (see TS09).

\section{The Monte-Carlo simulations}\label{MC}

The parameters of the simulations are $q$, $C_1/C_0$. We fix   value of $C_0\sim 0.01c$, $x_{out}=3$ and $q=0.01$ and we change
$\dot m$ from $0.01$ to $10$. As it is seen from Eq.  (\ref{temp_tau_0})   the plasma temperature $T_e$ of the corona region is not affected if  the disk cooling factor   $q\ll1$.    Therefore, for most of the simulated cases, we   fix the $q$ value at $0.01$. Once $q$ and $C_1/C_0$ are set, we compute  $\tau_{ff}$, $\tau_0$, and $T_e$ using Eqs.  (\ref{t_ff}),  (\ref{t_0-t_ff_new})  and (\ref{temp_tau_0}), respectively.

The physical model for the Monte-Carlo (MC)  simulations is  the Comptonization of the soft disk photons
in the converging flow of optical depth $\tau_{ff}$ extended from $r_{\rm S}$  to $ r_{out}=3r_{\rm S}$, surrounded by a spherical shell
of  optical depth $\tau_0$, corresponding to the transition layer.  The MC simulated spectra are not sensitive to the geometrical size of the TL layer, the radius of which being set to $10r_{\rm S}$, but is very sensitive to $\tau_0$, $\tau_{ff}$ and $T_e$.
Thus, we have  incorporated  plasma free-fall accretion  onto the central black hole in the converging flow region as  is described in  LT99.  Additionally, we include the thermal motion of the electrons in the converging flow and transition layer, simulated for   electron temperature $T_e$ given by
Eq. (\ref{temp_tau_0}).

 The seed X-ray photons were generated uniformly and isotropically
 at the surface of the border of the accretion disk, from $10r_{\rm S}$ to
 $12r_{\rm S}$. The soft seed photon spectrum is a blackbody spectrum with a temperature of 0.9 keV, which is close to the color BB temperature found in the spectra of  BHC sources
  [see e.g. \cite{BOR99}].

 In our MC simulations,  we  followed the trajectory of each photon
 in the following way:   first, a uniform deviate U is generated such that  $\tau_{ph} = -\ln U$,  $\tau_{ph}$ being the  optical thickness   a photon will
travel in curved geometry, we use the Schwarzschild metric, before scattering off an electron.
We then integrated the optical thickness  $\Delta \tau = n\sigma \Delta l$
along the photon path up to $\tau_{ph}$, taking into account the variation of the cloud number density $n $ and of the cross section $\sigma $ with the radius. Here the cross section  $\sigma $ is the Klein-Nishina cross section averaged over the local relativistic thermal electron
distribution.
 The gravitational red (or blue) shift endured by the photon was also computed at each
 step of this integration.

 We simulate  photon Compton scattering off an electron
 that includes
 the exact motion of the electron.  We also take into account the case where, at the end of the integration, the computed optical thickness has reached $\tau_{ph}$ and the photon has not left the considered volume [see \cite{POZ83}].
 In our procedure, we first compute the scattering electron momentum,
 and derive the scattered photon and electron characteristics
 from the Compton scattering kinetics.  Then we  check to determine if this event is
consistent with the Compton  scattering probabilities;
 if it is, then  the event is kept; if  it is not, another scattering photon is
generated, and the process goes on until the event is accepted.
 This process was successfully checked by comparing its results
 (see LT99)
 with the analytical results presented in \cite{tz98}.
 Once the new energy and direction of the photon has been determined by the
 Compton kinetics, we track it in the same way as above until it makes
 another scattering, it escapes from the considered volume, or until it is
``absorbed"   by the black hole at its horizon.

\section{ A combined effect of thermal and bulk motion Comptonization. BH spectral evolution}\label{thermal+bulk}

For a BH of the dimensionless spin $a<0.8$ the radius of the marginally stable orbit in the disk
$r_{mso}$ is about $3r_{\rm S}$ [see \cite{rw71} and \cite{ll75}].
Thus, below $r_{mso}$ and if the plasma temperature of the flow is essentially  non-relativistic, the accretion flow is almost in a free-fall regime, i.e.  advection dominated,
where the main effect of the up-scattering of the disk soft photons is the dynamical (or bulk) Comptonization \citep{LT99}.

\begin{figure}[ptbptbptb]
\includegraphics[scale=1.0, angle=0]{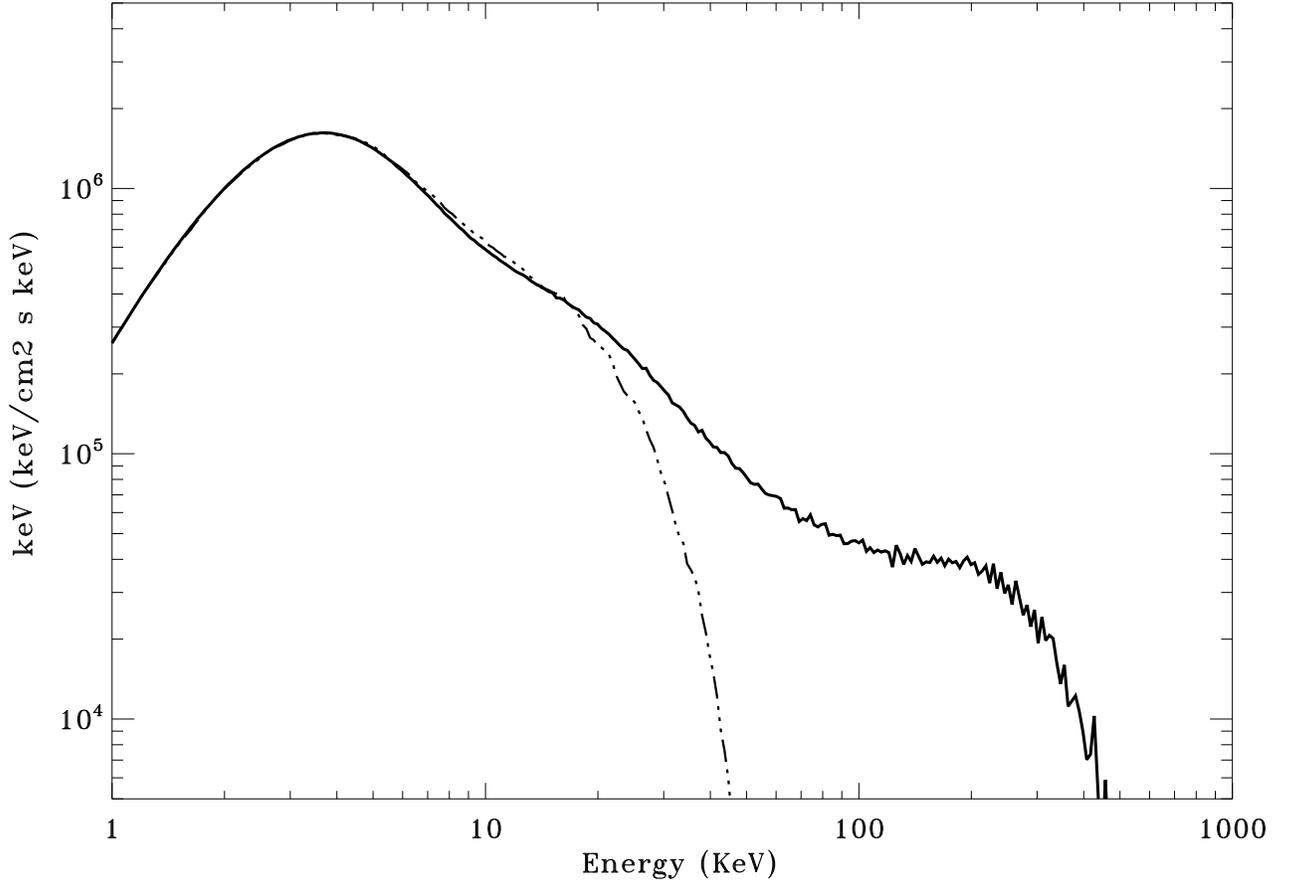}
\caption{The appearance of  Bulk Motion Comptonization in the Monte-Carlo simulated  spectrum.
The spectrum is shown in units of keV (keV cm$^{-2}$s$^{-1}$keV$^{-1}$) with absolute values that are  multiplied by an arbitrary constant.
The dotted line corresponds to a simulation with an assumption that there is no bulk effect in the converging flow zone. The solid line results from the same simulation adding bulk velocity. The effect of bulk motion Comptonization is clearly seen above 20-30 keV.   The parameters used are: mass accretion rate in Eddington units $\dot m=5$,   flow electron temperature  $kT_e=4.3$ keV, $C_1/C_0=7$ and $q=0.01$. }
\label{fig3}
\end{figure}

In Figure \ref{fig3}, we illustrate the effect of bulk motion Comptonization on the spectral signature.
The parameters for the simulated spectrum were $\dot m=5$, $q = 0.01$, and $C_1/C_0 = 7$, resulting in and inferred  electron temperature of the flow  $kT_e=4.3$ keV
(see Eqs. \ref{temp_tau_0},  \ref{t_ff}, \ref{t_0-t_ff_new}).  The dotted line corresponds to the simulated spectrum where the bulk effect of the converging flow is not taken into account and  the solid line corresponds to the spectrum  where the bulk velocity of the converging flow is included in the simulations.
The effect of bulk (dynamical Comptonization) flow  is clearly seen above 20-30 keV. The steep power-law component that emerges in the spectrum which is now extended up to 200 keV.

All our simulated spectra can be represented by an additive XSPEC model consisting of the addition of two, so called ``bulk motion'' Comptonization (BMC) components: a BMC with a high energy cut-off ({\it BMC1} component) and a relatively soft  {\it BMC2} component with a high energy cut-off. Thus, our model for fitting is $(bmc1*highecut1+bmc2*highecut2$). The {\it BMC} model describes the outgoing spectrum as a convolution  of the input ``seed'' blackbody spectrum, with a normalization  of
$N_{bmc}$ and color temperature is $kT$,  with the  Comptonization Green's function which is a broken power law of spectral index $\alpha$ (photon index $\Gamma=\alpha+1$).
The resulting spectrum is characterized by the parameter $\log(A)$ related
to the Comptonized fraction $f$ as $f=A/(1+A)$ and  spectral index
$\alpha$.

\begin{figure}[ptbptbptb]
\includegraphics[scale=0.7,angle=-90]{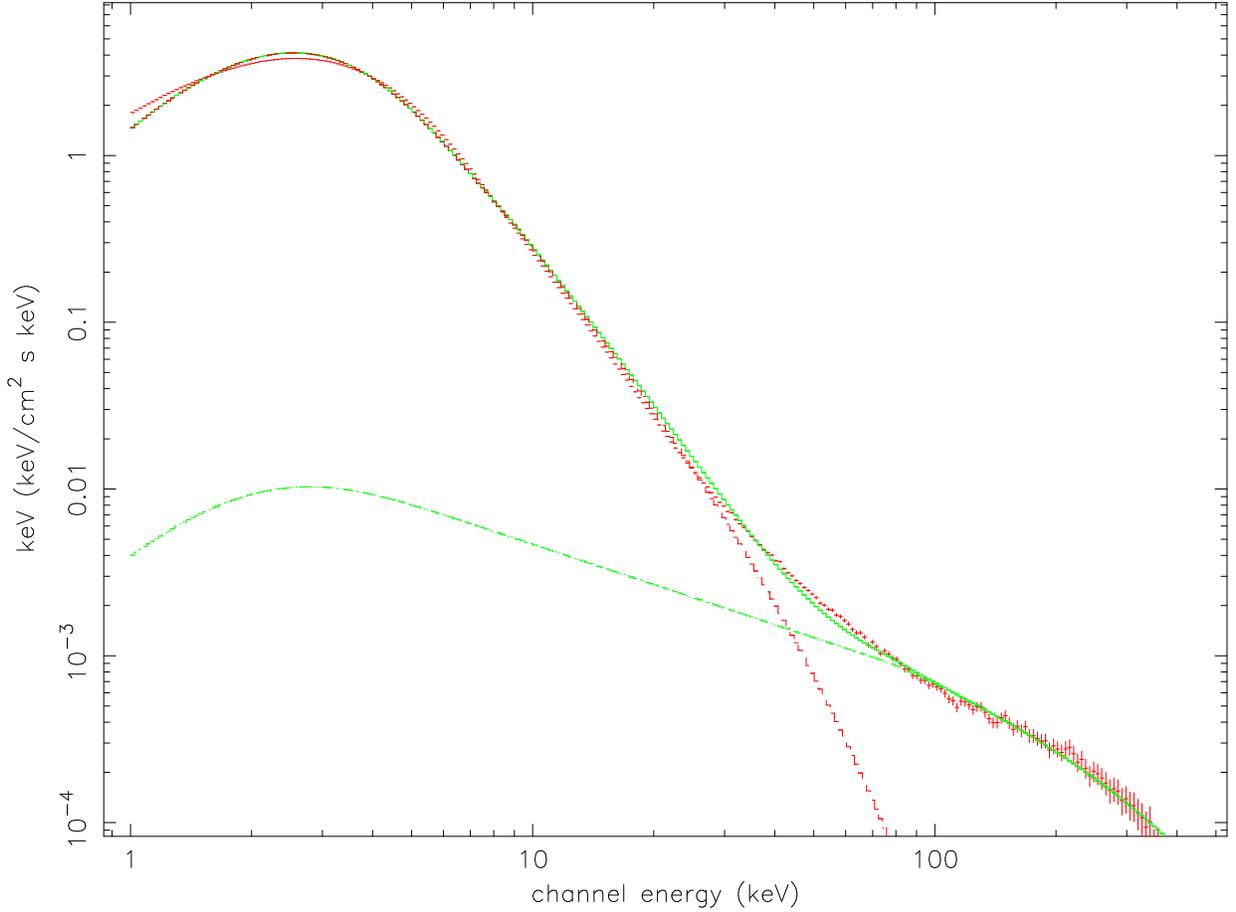}
\caption{ An example of simulated bulk+thermal Comptonization spectrum (red histogram)  and the XSPEC  model {\it bmc1$\ast$highcut1+bmc2$\ast$ highcut2}.
The spectrum is shown in units of keV (keV cm$^{-2}$s$^{-1}$keV$^{-1}$) with absolute values that are multiplied by an arbitrary constant.
 The hard  and soft components are shown by dash green and  dash red curves, respectively. The resulting spectrum is shown by a solid green curve.    In the simulations, the  mass accretion rate in Eddington units is $\dot m=5$, $C_1/C_0=5$, $q=0.01$ and the  flow electron temperature is $kT_e=2.4$ keV.}
\label{fig4}
\end{figure}

In Figure \ref{fig4} we show an example of the simulated spectra produced by two BMC components with high energy cutoff, specifically   {\it bmc1$\ast$highcut1+bmc2$\ast$ highcut}2.
The spectrum is shown in units of keV (keV cm$^{-2}$s$^{-1}$keV$^{-1}$) where absolute values  are multiplied by an arbitrary constant.
 The hard component is given in dash green and  the soft component is given in dash red. The resulting best-fit spectrum is presented by the solid green curve.    In our simulations the mass accretion rate in Eddington units, is $\dot m=5$, $C_1/C_0=5$, $q=0.01$ and  the flow electron temperature is $kT_e=2.4$ keV. The characteristics (parameters) of $bmc1$ and $highcut1$ are  spectral index
 $\alpha_h=1.8$ ($\Gamma_h=2.8$),  $\log A=0.26$,  cutoff energy $E_{cut}=100$ keV, efold energy $E_{fold}=500$ keV and that for $bmc2$ and $highcut2$ are $\alpha_{sf}=2.23$ ($\Gamma_{sf}=3.33$),
 $\log A=0.99$,   $E_{cut}=4.5$ keV,  $E_{fold}=4.2$ keV.

The general picture of LHS-IS-HSS transition is illustrated
in Figure \ref{fig5}  where
we bring together simulated spectra related to the observational spectra of  LHS, IS, HSS and VSS  to demonstrate the BH spectral evolution from the  low-hard to soft states.
We reproduce the observational spectral evolution by varying the mass accretion rate $\dot m$  from
0.05 to 5.  Top and bottom panels correspond to   $C_1/C_0=5$ and  $C_1/C_0=10$,
respectively.
The different shapes of the simulated spectra are related to  the different spectral states in the following manner: low $\dot m=0.05$ (black histogram), which the TL optical depth  $\tau_0= 2.5,~2$  and $kT_e=40, ~55$ keV for $C_1/C_0=5$ and
 $10$, respectively (see Fig. \ref{fig2}), is similar to the observed   LHS spectrum (see ST09, TS09); increasing the accretion rate to $\dot m=0.1$ (blue histogram), the spectrum becomes softer (i.e. power law is steeper and exponential cutoff energy is shifted to lower energies  than that for  $\dot m=0.05$) and the Comptonization component dominates the blackbody (BB) component, which is barely seen  in  3-150 keV energy range is similar to the spectrum seen in the beginning of the IS state; increasing $\dot m$ further produces a spectra (green histogram)   characterized by a strong soft BB component  and an extended steep power law, which is comparable to the observed HSS spectra.

 For moderate $\dot m=0.5$  (red histogram) and $C_1/C_0=5$ the spectral shape is a blackbody-like but slightly modified by the thermal Comptonization because $kT_e\sim 4$ keV  is very low for  efficient thermal up-scattering and $\tau_{ff}\sim 1.6,  \dot m= 0.8$ is also low for the bulk motion Comptonization.
A similar spectra is detected during the very soft state (VSS).
However  when mass accretion further increases to a value of $\dot m=5$ the extended high energy component is seen up to energies 300 keV  as a result of the bulk motion Comptonization in the converging inflow, $r<r_{\rm S}$.    

For  $C_1/C_0=10$ and  $\dot m >1$ the spectral shapes  are strongly affected by thermal  Comptonization because $kT_e\sim15$ keV and $\tau_0\sim 8-10$ are relatively high  [see Figs. \ref{fig2},  \ref{fig5}  (lower panel)].  
One can also observe  the effect of bulk motion Comptonization   seen as an extension of  the emergent spectrum to energies of 300 keV when mass accretion rate increases,   see green curve  for 
$\dot m=5$  and compare the similar effect  for   $C_1/C_0=5$
(Fig. \ref{fig5},  lower  and upper panels respectively).

\section{Indices of soft and hard components of the X-ray simulated spectra as a function of   mass accretion rate}
\label{MC_index-mdot}
We have made  series of simulations with different values of  the $C_1/C_0$ ratio in order to study the evolution of $C_1/C_0$ - index saturation. This evolution is shown on Figures  \ref{fig6}-\ref{fig8},
 where the saturation limits for  $\Gamma_{sat, s}$ and $\Gamma_{sat, h}$ are shown.
 Moreover, we show the observed pattern of index vs $\dot m$ (see ST09 and TS09) can be reproduced in our simulations by the variation of only one parameter of the velocity profile, i.e. $C_1/C_0$. In Figures \ref{fig6}-\ref{fig8}, we show  $\Gamma_{sf}$ (red points) and
$\Gamma_h$ (blue points) as a function of $\dot m$ for a given $C_1/C_0$.  Saturation of $\Gamma_{sf}$ occurs in all simulated cases. The highest level saturation
 of $\Gamma_{sf}$  is achieved at $C_1/C_0 =5$, when the matter accumulation  is strong in the transition layer (see Fig. \ref{fig9}). The saturation value $\Gamma_{sat, sf}\sim 4.9$ is close to the  saturation index  value (about 4.3)  of the soft component of {\it RXTE} X-ray spectrum  from GRS 1915+105 (TS09).
In \S \ref{model}, we argue that  index saturations of the soft component
at high $\dot m$ is expected for each given  value of $C_1/C_0$.
When  $V_{MA} =0.01(1+C_1/C_0)c$ is low  (or $C_1/C_0\lax 5$),
the TL layer optical  depth $\tau_0$ is  high (see   Eq. \ref{t_0-t_ff_new}) and consequently the plasma temperature of accretion flow   $T_e$ is low  (see Eq. \ref{temp_tau_0}). 
In other words matter accumulation efficiency in the transition layer (or $\tau_0$) is inversely proportional to magneto-acoustic velocity   $V_{MA}$ there (see Eq. \ref{t_0-t_ff}).

ST09 and TS09 present  strong observational and theoretical arguments that the saturation of  index $\Gamma_h$ occurs at  high $\dot m$ too.  Specifically,  ST09 argues that, in the converging flow,  the spectral index,  as an  inverse of Comptonization $Y-$ parameter,  $\alpha\sim1/Y=1/(\eta N_{sc})$, should saturate  at high  $\dot m$  because  a linear increase of $N_{sc}$ with   $\dot m$ is compensated for by a decrease of $\eta$ as $1/\dot m$.  Here $\eta$ is average fractional energy per scattering. 
Our simulations show that the index saturation values
of  $\Gamma_{sat, h}$  weakly depend on $C_1/C_0$ ratio i.e. $\Gamma_{sat, h}$ is in the range of 2.6-2.8 for all $C_1/C_0\gax 5$ (see Fig. \ref{fig9}).

Particular for $C_1/C_0=5$,  the hard component of the simulated spectrum (above 80 keV) is  absent 
when  $\dot m$ increases from 0.1 to 0.5.
(see Fig. \ref{fig5}). 
However one can see  that the spectral hard component  arises when $\dot m$ increases above 1 (see also Fig. \ref{fig6}). 

Moreover,  for $C_1/C_0=5.5$,  and   $\dot m\gax 0.5$  the flow electron temperature $T_e$   decreases to about  3 keV (see Fig. \ref{fig2}). In this case,
we obtain that  $\Gamma_h$  and $\Gamma_{sf}$ saturate to $\sim2.7$ and $\sim4.3$, respectively (see Fig. \ref{fig9}). This situation is probably realized in the innermost part of  GRS 1915+105 where there  are similar values of   $\Gamma_h\sim 2.7$  and $\Gamma_{sf}\sim 4.3$ inferred from  {\it RXTE} X-ray spectrum (see TS09). 

  One can conclude that  for TL velocities $V_{MA}$ of order $0.05c$ and less, when the accumulation effect in  the transition layer is very strong, the steep extended power law, as a  spectral signature of the converging, is screened by the optically thick material of the transition layer.
In this case, the end of the spectral evolution from  the low-hard state is the very soft state
characterized by a strong blackbody like component
(see the  red and green simulated spectra in the upper panel of Fig. \ref{fig5}).
The soft spectral component  is dominant  at  $\dot m\geq0.5$ and an extended hard tail is  only seen in the spectrum for high values of $\dot m$ (see green spectral  histogram in Fig. \ref{fig5}).  TS09 show that this type of spectral evolution is  seen in {\it RXTE} observations of GRS 1915+105.

However, for $V_{MA}\gax 0.09c$, the accumulation effect is not as strong in the TL 
and as a consequence the steep power law extended up to 300 keV is seen at 
higher level of count rate then that for $V_{MA}\sim 0.05c$ (compare the upper and lower panels in Fig. \ref{fig5}).
In this case, we see a classical spectral evolution from LHS to VHS through the intermediate state that is observed in many BH binaries  [see e.g \cite{rm}, ST09].

We reproduce, using our Monte Carlo simulations,  the observed spectral evolution in BH binaries  when we assume  the velocity profile, approximately constant in the transition layer and followed by free-fall below $3R_{\rm S}$. In fact, because the transition layer shrinks with mass accretion rate 
[see \cite{tlm98}] one can assume the constancy of $V_{MA}$ there. Thus  we proceed with our Monte-Carlo simulations of photon propagation in this kind of atmosphere.

However in the framework of our spectral evolution  model, we do not obtain an extended power law  with the index saturation of 2.1 that has been detected in HSS of Cyg X-1, GX 339-4, 4U 1543-47 (see ST09). This kind of the hard tail is reproduced in our  simulations when we assume that there is a sub-Keplerian, relatively hot accretion flow  with the same temperature as the transition layer, in addition to the disk accretion flow. In these simulations, the ratio of the sub-Keplerian to the Keplerian accretion rate is $\dot m_{sub}/\dot m_{\rm K}=50$ (see Fig. \ref{fig10}). This idea of  advection dominated accretion flow (ADAF), first introduced
by  \cite{ny94}, is very popular in  the astrophysical community.

To illustrate this, we use Cyg X-1, where it has been shown [see e.g. \cite{p78}, \cite{k98}]  that the X-ray source is powered mainly by accretion from the strong
stellar wind of the supergiant star. There is a  strong observational indication of the presence of sub-Keplerian flow in this source. The temperature of this flow $T_{e,sub}$ can be dictated  by the photons Comptonized in the transition layer. Presumably when the disk mass accretion $\dot m_{\rm K}$ increases
the sub-Keplerian  mass accretion rate $\dot m_{sub}$ increases too. In order to have the relatively
hot ADAF, one varies $\dot m_{\rm K}$ in the low value range.
To show this, the variation of $\dot m_{\rm K}$ from $0.05$ to $0.1$   leads  to a decrease of
 $T_e\sim T_{e,sub}$   from 60 to  20 keV (see Fig. \ref{fig2}).  
  \cite{LT99} obtained that the photon index of the converging flow  power-law tail $\Gamma_h$ saturates to $\sim 2.1$ when   $T_{e,sub}\sim 30$ keV. 
  
  Hence we obtain the observable saturation index of 2.1   assuming the existence of two component flows in the aforementioned BHs.

\section{Evolution of spectral efold energy  $E_{fold}$ with mass accretion rate $\dot m$}
\label{E_cf-mdot}

Dynamical (bulk)  Comptonization spectra can be distinguished from thermal Comptonization spectra when there is information about the evolution of $E_{fold}$ with $\dot m$.  This information is inferred from X-ray  observations of spectral transitions in BHCs, where the evolution  patterns  are drastically different between the two Comptonizations.
In Figure \ref{fig11}, we plot how the high energy efold $E_{fold}$ energy varies  with the mass accretion rate $\dot m$.  In the {\it bulk motion Comptonization} case  
the efold energy $E_{fold}$ decreases and then increases when  mass accretion rate $\dot m\gax 1$ (see   Figure \ref{fig11}).

Note that in the thermal Comptonization spectra $E_{fold}$  is proportional  to the plasma temperature $T_e$ ($E_{fold}\sim 2 kT_e$).
Given that $kT_e$ saturates to  constant  value of  2 keV  for $\dot m\gax 1$ (see Fig.  \ref{fig2}), $E_{fold}-$energy  should  also  saturate to  low values of  $E_{fold}\sim 3-4$ keV  
[see \cite{ft10}  for  a similar situation in the  neutron star case]. 


\section{Conclusions}\label{summary}
\begin{figure}[ptbptbptb]
\includegraphics[scale=0.7,angle=0]{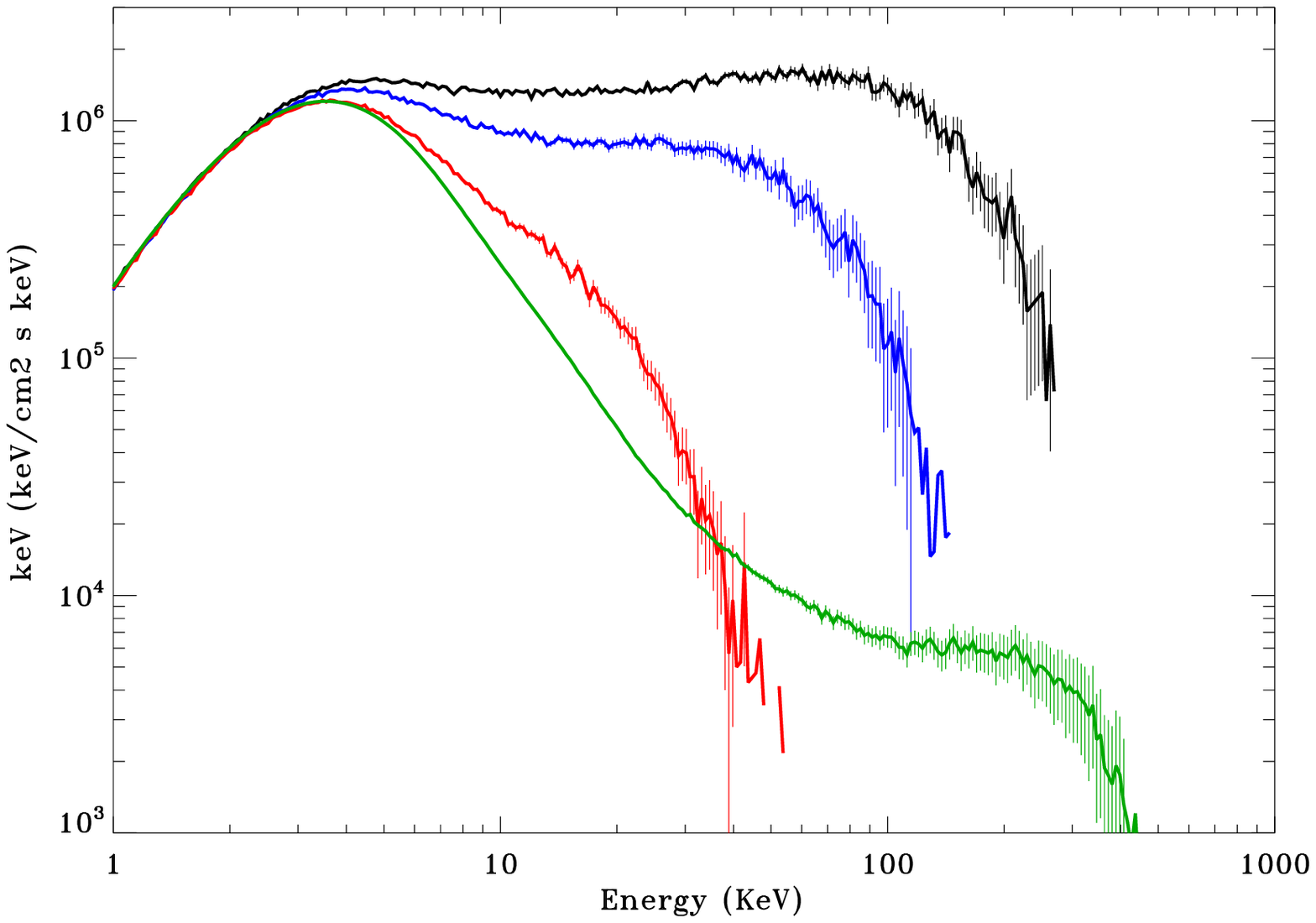}
\includegraphics[scale=0.7,angle=0]{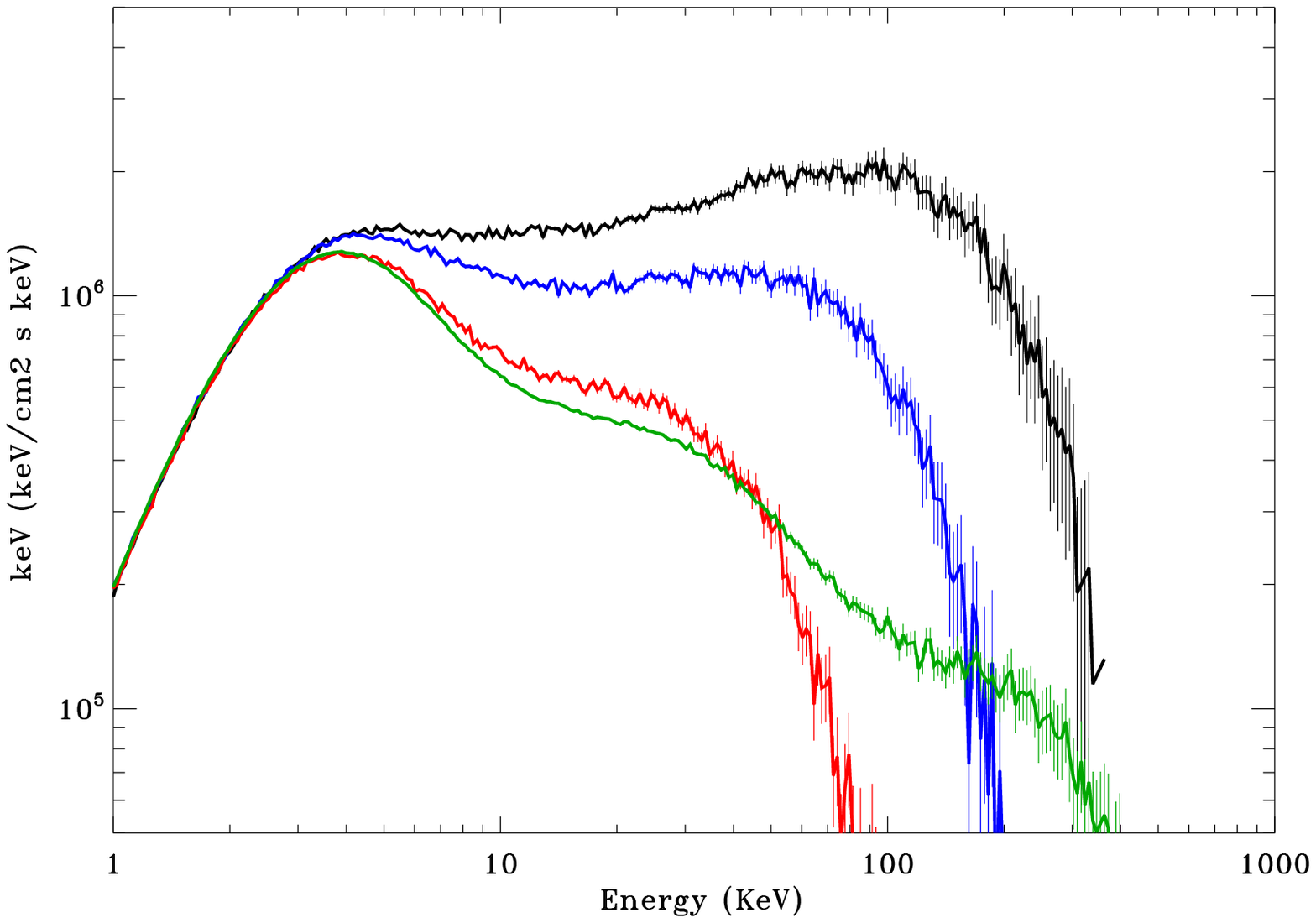}
\caption{Spectral evolution as a function of mass accretion rate:
upper panel:   $C_1/C_0=5$, bottom panel:  for $C_1/C_0=10$. Black, blue,  red and green histograms correspond to $\dot m=0.05, ~0.1,~ 0.5$ and 5, respectively.
}
\label{fig5}
\end{figure}

\begin{figure}[ptbptbptb]
\includegraphics[scale=0.7,angle=0]{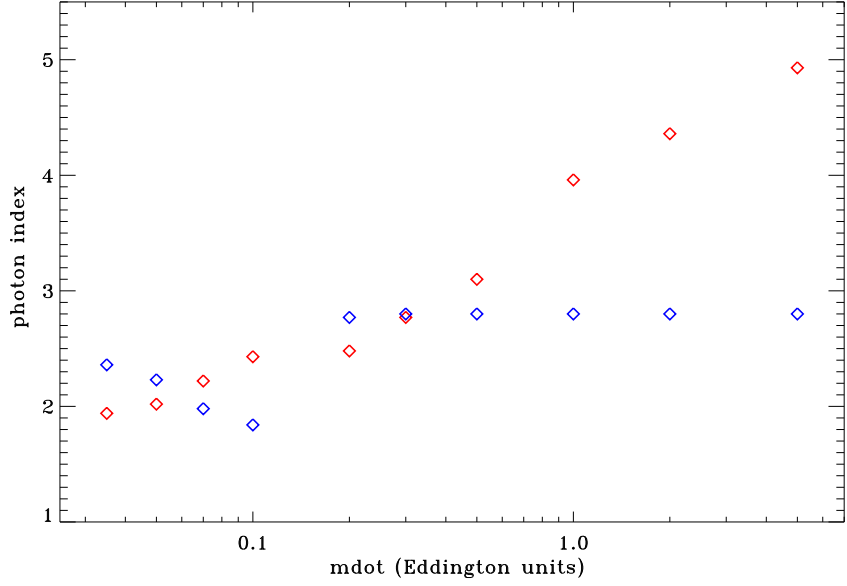}
\caption{Index vs $\dot m$  for two BMC components:
red points are for the soft component and blue points are for the hard BMC component C1/C0=5.}
\label{fig6}
\end{figure}

\begin{figure}[ptbptbptb]
\includegraphics[scale=0.7,angle=0]{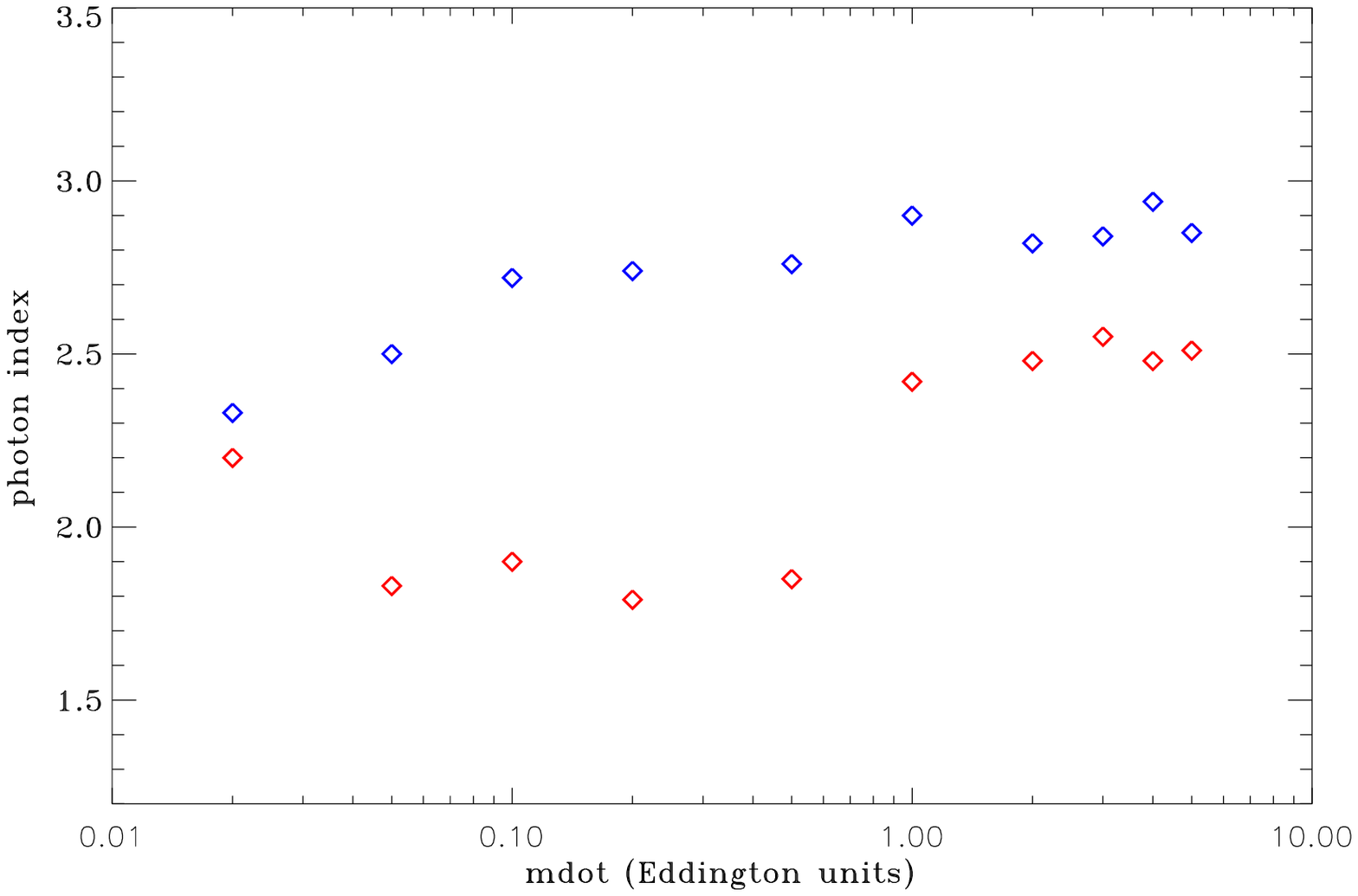}
\caption{ Same as that in Figure \ref{fig6} but for  C1/C0=8.}
\label{fig7}
\end{figure}

\begin{figure}[ptbptbptb]
\includegraphics[scale=0.7,angle=0]{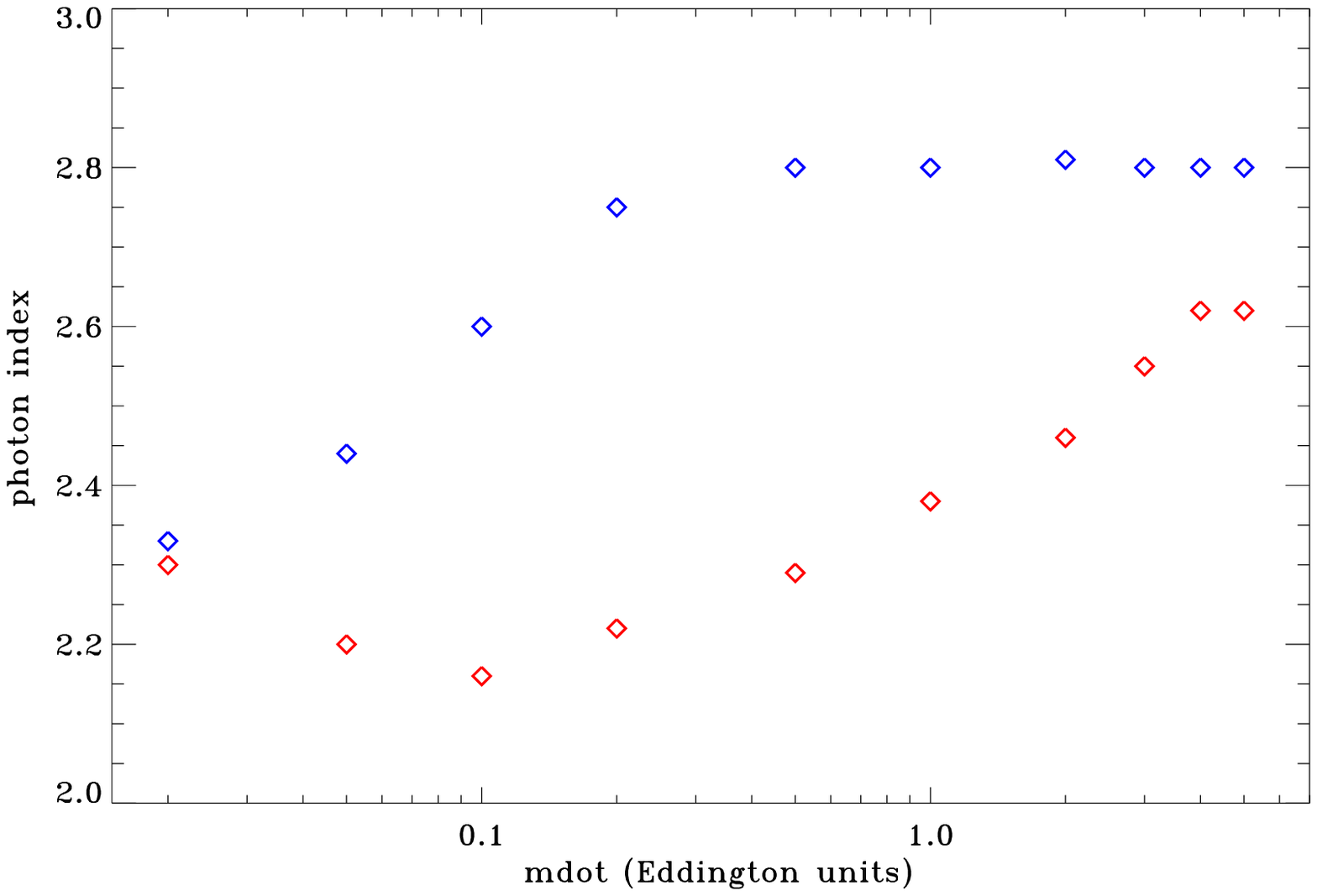}
\caption{ Same as that in Figure \ref{fig6} but for  C1/C0=10.}
\label{fig8}
\end{figure}

\begin{figure}[ptbptbptb]
\includegraphics[scale=0.7,angle=0]{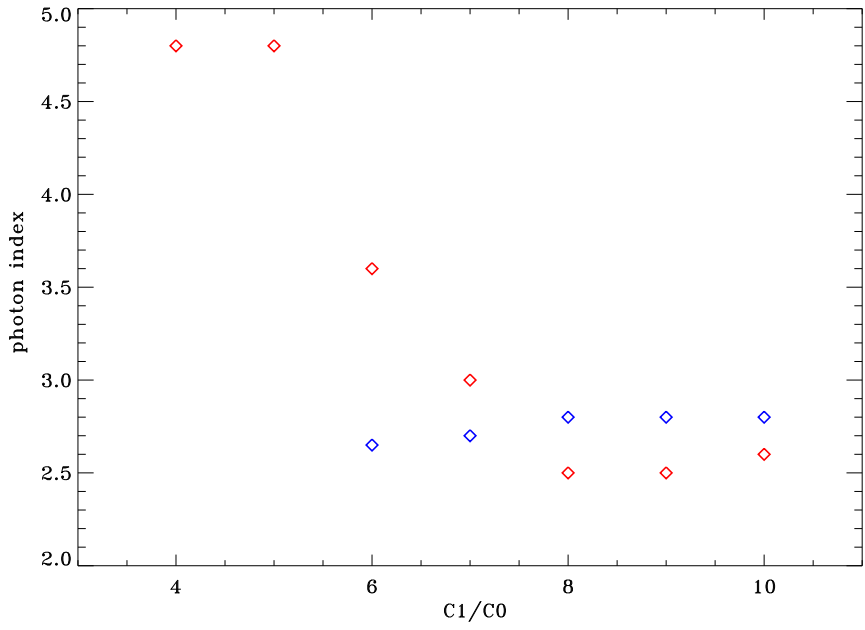}
\caption{Saturation values of index as a function of C1/C0. }
\label{fig9}
\end{figure}

\begin{figure}[ptbptbptb]
\includegraphics[scale=0.7,angle=0]{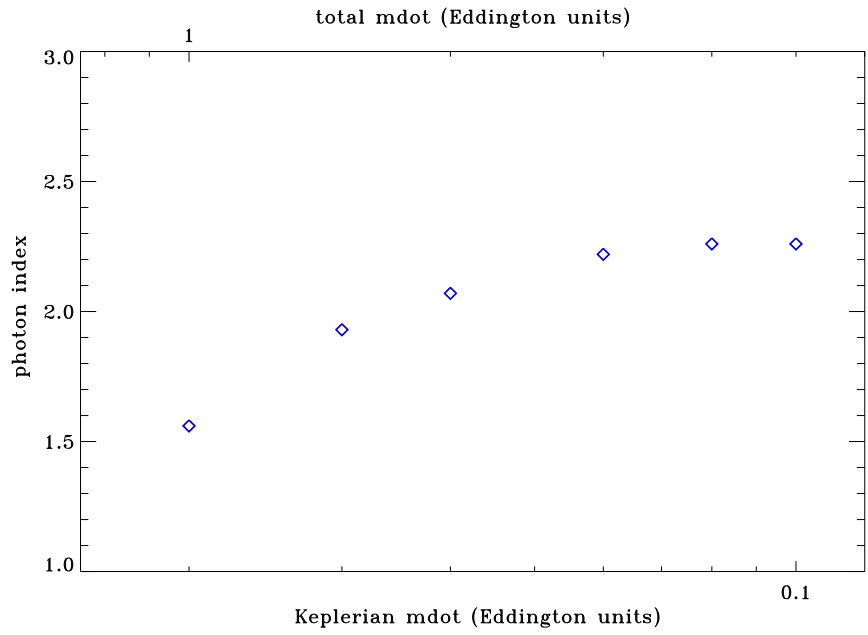}
\caption{Index  vs $ \dot m$ in the presence of sub-Keplerian flow above the disk.
The accretion flow has the same electron temperature as the transition layer one. The ratio of $\dot m_{sub}/\dot m_{\rm K}=50$. Total $\dot m_{tot}=\dot m_{\rm K}+\dot m_{sub}$}
\label{fig10}
\end{figure}

\begin{figure}[ptbptbptb]
\includegraphics[width=4.8in,height=2.75in, angle=0]{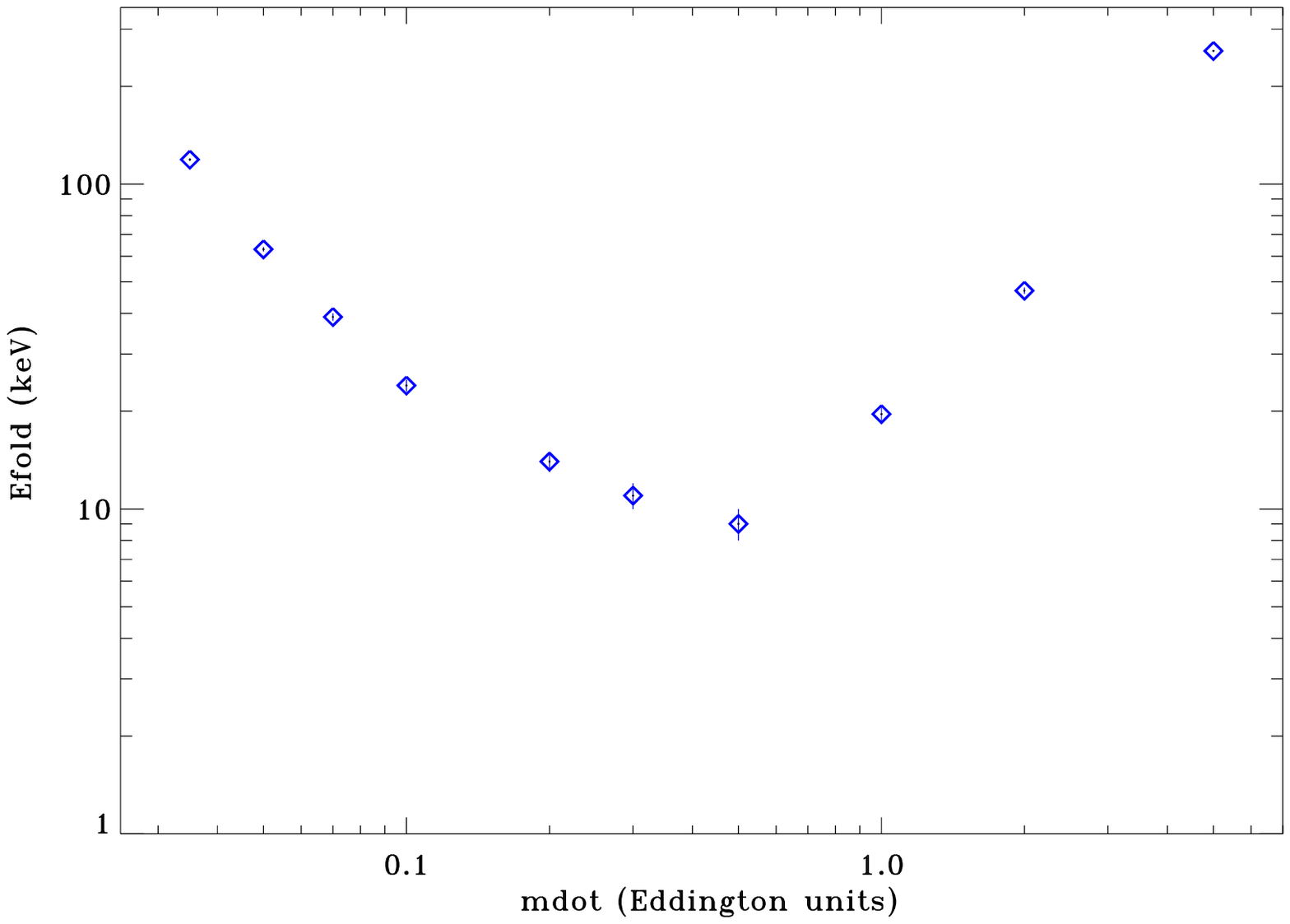}
\includegraphics[width=4.8in,height=2.75in, angle=0]{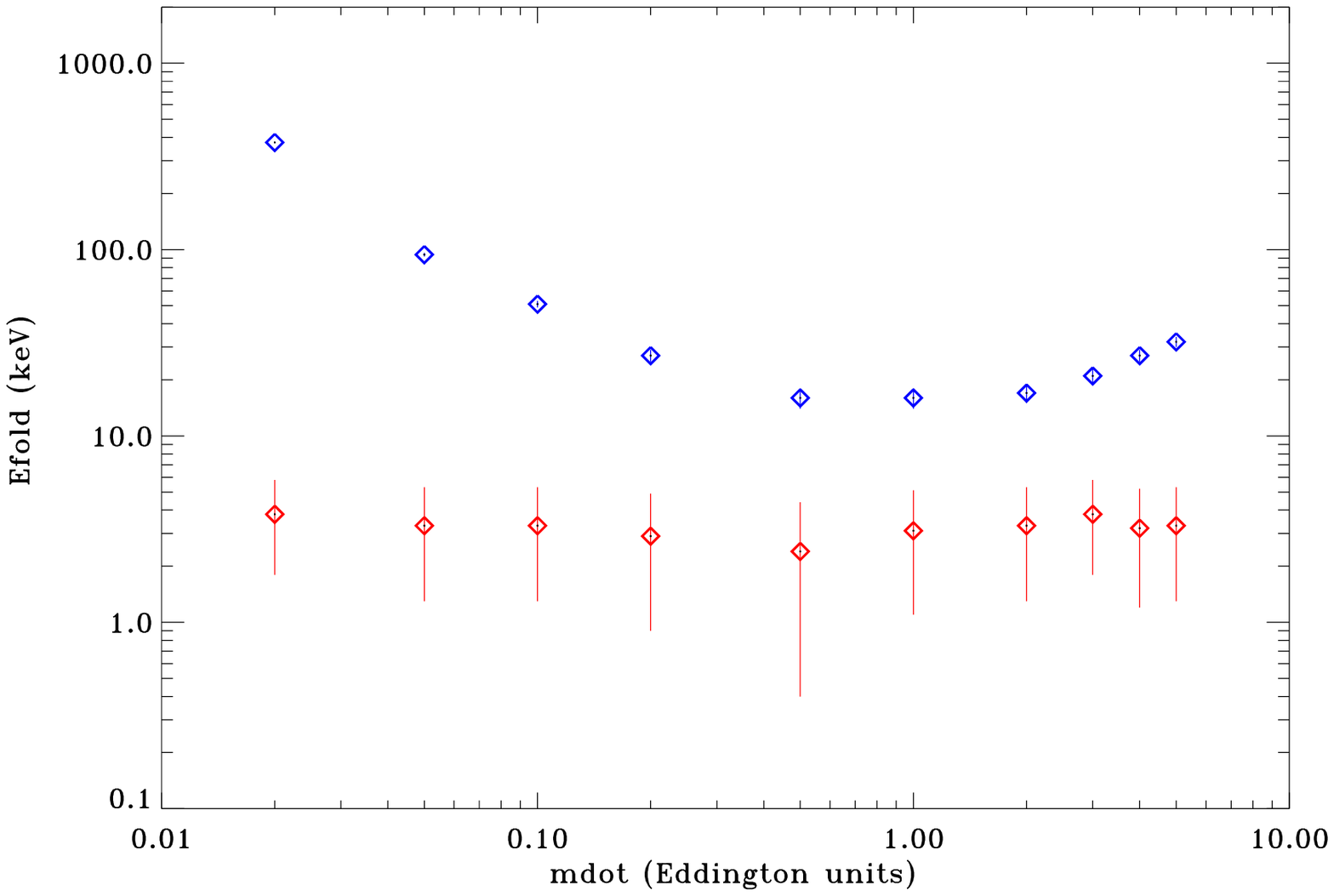}
\includegraphics[width=4.8in,height=2.75in, angle=0]{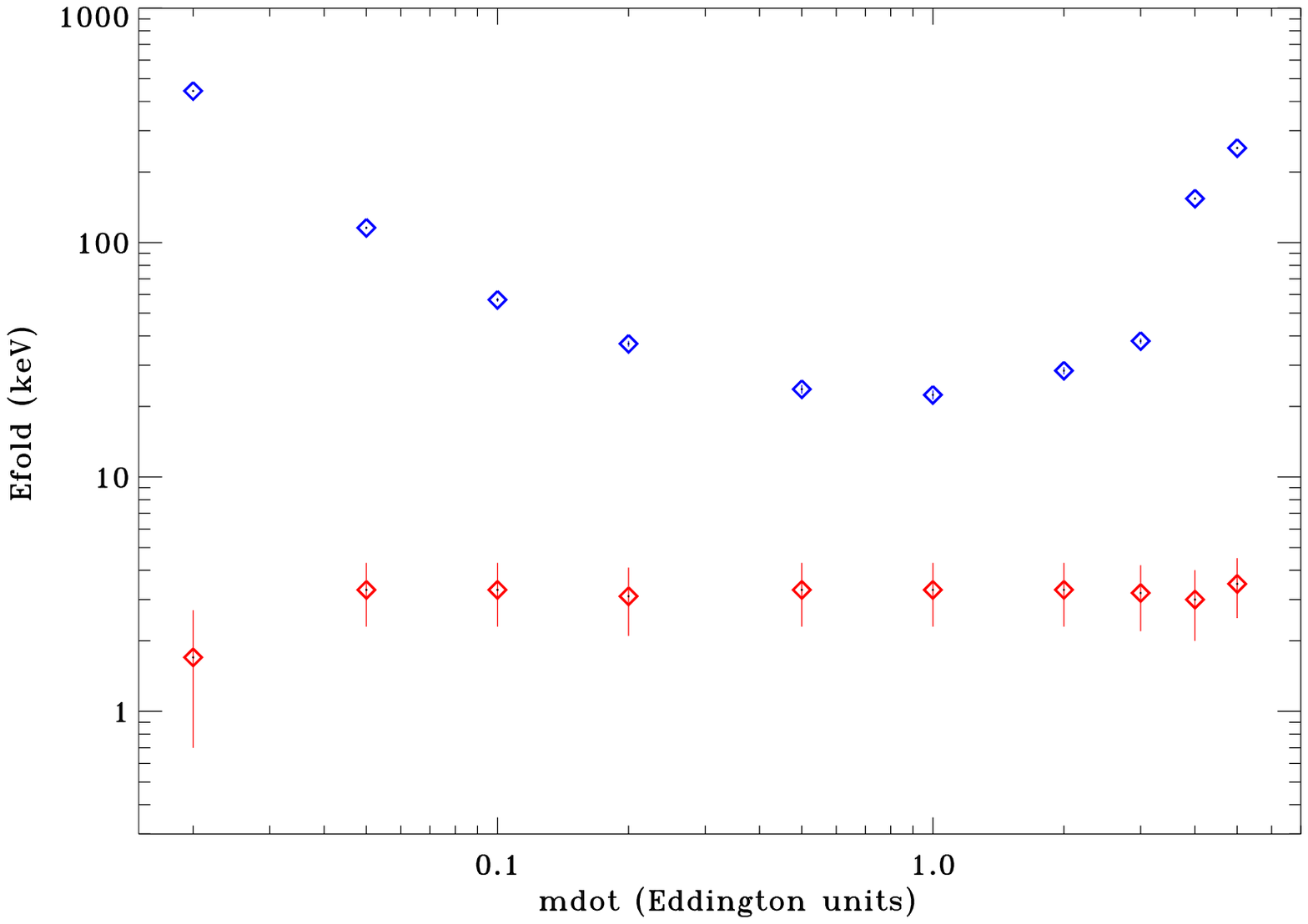}
\caption{$E_{fold}$ energy vs $\dot m$:
 $C_1/C_0=5,~8,~10$ upper, middle and bottom panels respectively.}
\label{fig11}
\end{figure}

We have studied, using Monte Carlo simulations, the evolution of X-ray energy spectra from the low-hard state to the high-soft state, and found that a hard spectrum should be observed at relatively low values of mass accretion rate independently of the velocity profile (see Fig. \ref{fig5}). Further, for  low velocities of the accretion flow in the transition layer (TL)  $v\sim 0.05c$,  there is plasma  accumulation at high mass accretion rates that produces high optical depth and thus one  can barely see  the high energy tail. 
 An emergent spectrum  is characterized by a blackbody like  shape of temperature on order of 1 keV and the relatively weak hard tail (see the green spectral histogram in the  upper panel in Fig. \ref{fig5}). 
This kind of spectra is observed in the very soft state of BH binaries.
We also found that when the TL velocities  are of order of (0.07-0.1)c at high mass accretion rate  the extended power-law  component is more pronounced in the emergent spectra 
(see the green spectral histogram in the lower panel of Fig. \ref{fig5}). For all velocity values in TL,  our results predict index saturation vs mass accretion rate (see Figs. \ref{fig7}-\ref{fig8}).
Finally, we showed that the efold energy  $E_{fold}$ of the emergent spectra decreases and then  increases as the mass accretion rate increases (see Fig. \ref{fig11}). This is a specific property of BH spectra. This type of  behavior  of  $E_{fold}$ vs $\dot m$
(or vs photon index) has been recently discovered  by \cite{ts10}  in {\it RXTE} observations of  XTE J1550-564.

Note the decrease of the spectral efold energy vs mass accretion rate   occurs when the high energy photons  formed in the converging flow are not seen in the resulting spectrum. In this case, the spectral shape is only determined by  thermal Comptonization taking place  in the outer parts of Compton cloud. 

We appreciate the referee for careful reading  and his/her valuable  comments on the content of the manuscript. We also acknowledge the manuscript editing by Charlie Bradshaw.  

\clearpage
\newpage

\newpage

\newpage

\newpage

\newpage

\newpage

\newpage

\newpage

\clearpage
\newpage

\clearpage
\newpage

\newpage

\clearpage


\begin{thebibliography}{}
\bibitem[Belloni(2005)]{bell05} Belloni, T.\ 2005, Interacting
Binaries: Accretion, Evolution, and Outcomes, AIP Conference, 797, 197 (astro-ph/0504185)

\bibitem[Belloni et al. (2000)]{bell00} Belloni, T., Klein-Wolt, M., M{\'e}ndez., M.,  van der Klis, M.,
\& Paradijs, J. 2000,  \aap, 355, 271


\bibitem[Borozdin et al. (1999)]{BOR99}
Borozdin, K., Revnivtsev, M., Trudolyubov, S., Shrader, C. \& Titarchuk, L.   1999,
ApJ, 517, 367

\bibitem[Chakrabarti \& Titarchuk (1995)]{ct95}
Chakrabarti, S.K. \& Titarchuk, L. 1995, ApJ, 455, 623

\bibitem[Farinelli  \& Titarchuk (2010)]{ft10}
Farinelli,  R. \& Titarchuk, L.  2010, A\&A, in press

\bibitem[Kaper  (1998)]{k98}
Kaper, L.  1998,  in ASP Conf. Ser. 131, Boulder-Munich II: Properties of Hot,
Luminous Stars, ed. I. D. Howard (San Francisco: ASP), 427

\bibitem[Klein-Wolt  \& van der Klis(2008)]{kw08} Klein-Wolt, M., \& van der Klis, M.\ 2008, \apj, 675, 1407

 \bibitem[Landau \& Lifshitz (1975 )]{ll75} Landau, L.D. \& Lifshitz, E.M. 1975, The classical theory of fields (Pergamon Press: Oxford)


\bibitem[Laurent \& Titarchuk(2001)]{lt01}
Laurent, P., \& Titarchuk, L. 2001, ApJ, 562, 67

\bibitem[Laurent \& Titarchuk(1999)]{LT99}
Laurent, P., \& Titarchuk, L. 1999, ApJ, 511, 289 (LT99)

\bibitem[Narayan \& Yi  (1994)]{ny94}
Narayan, R \& Yi, I.  1994,  \apj,  428, L13

\bibitem[Nied\'{z}wiecki \&  Zdziarski  (2006)]{nz06}
Nied\'{z}wiecki, A. \&  Zdziarski, A. 2006, MNRAS, 365, 606

\bibitem[Peterson (1978)]{p78}
Peterson, K.  1978,  \apj,  224, 625

\bibitem[Pozdnyakov et al  (1983)]{POZ83}
Pozdnyakov, L.A., Sobol', I.M.  \& Sunyaev R.A. 1983, Astrophys. Space. Phys. Rev., 9, 1

\bibitem[Ruffini \& Wheeler  (1971)]{rw71} Ruffini, R.,  \& Wheeler, J.~A.\  1971, Physics Today, 24, 1, 30

\bibitem[Remillard \& McClintock (2006)]{rm} Remillard,
R.~A., \& McClintock, J.~E.\ 2006, \araa, 44, 49  (RM06)

\bibitem[Seifina \& Titarchuk (2010)]{st10}
Seifina, E.  \& Titarchuk, L.     2010, \apj,  721,  , (ST10)

\bibitem[Shakura \& Sunyaev  (1973)]{ss73}
Shakura, N.I. \& Sunyaev, R.A.   1973,  A\&A, 24, 337

\bibitem[Shaposhnikov \& Titarchuk (2009)]{st09}
Shaposhnikov, N. \& Titarchuk, L.    2009, \apj, 699, 453

\bibitem[Shaposhnikov \& Titarchuk (2006)]{ST06}
Shaposhnikov, N., \& Titarchuk, L. 2006, \apj, 643, 1098

\bibitem[Sunyaev \& Titarchuk (1980)]{ST80}
Sunyaev, R.A. \& Titarchuk, L.G.  1980,  A\&A, 86, 121 (ST80)

\bibitem[Titarchuk, Bradshaw \& Wood (2001)]{tbw01}
Titarchuk, L. \& Bradshaw, C.  \& Wood, K.   2001, \apj,  560, L58


\bibitem[Titarchuk, \& Fiorito (2004)]{tf04}
Titarchuk, L.G. \& Fiorito, R.  2004,  \apj, 612,  988 (TF04)

\bibitem[Titarchuk et al.  (1998)]{tlm98}
Titarchuk, L., Lapidus, I. \& Muslimov, A.   1998,  \apj, 499,  315 

\bibitem[Titarchuk, Mastichiadis \& Kylafis (1997)]{tmk97}  Titarchuk, L., Mastichiadis, A., \& Kylafis, N. D., 1997,  \apj, 487, 834

\bibitem[Titarchuk \& Seifina (2009)]{ts09}
Titarchuk, L. \& Seifina, E.    2009, \apj,  706, 1463 (TS09)

\bibitem[Titarchuk \& Shaposhnikov  (2010)]{ts10}
Titarchuk, L. \&  Shaposhnikov, N.   2010, \apj,  in press

\bibitem[Titarchuk \& Shaposhnikov  (2008)]{ts08}
Titarchuk, L. \&  Shaposhnikov, N.   2008, \apj,  678, 1230

\bibitem[Titarchuk \& Shaposhnikov (2005)]{ts05}
Titarchuk, L., \&  Shaposhnikov, N. 2005, ApJ,  626, 298

\bibitem[Titarchuk, Shaposhnikov \& Arefiev  (2007)]{tsa07}
Titarchuk, L.,  Shaposhnikov, N. \& Arefiev, V.  2007, \apj,  660, 556

\bibitem[Titarchuk \& Zannias (1998)]{tz98}
Titarchuk, L., \& Zannias. T. 1998, ApJ, 493, 863

\bibitem[Vignarca et al. (2003)]{vg03}
Vignarca, F., Migliari, S., Belloni, T., Psaltis, D., \& van der Klis, M. 2003,  A\&A, 397, 729


\end{thebibliography}
\end{document}